\documentclass[a4paper,10pt]{article}
\usepackage{amsthm}
\usepackage{mathrsfs}
\usepackage{amsfonts}
\usepackage{color}
\usepackage{graphicx}
\usepackage{amsmath}
\usepackage{amsfonts}
\newtheorem{theorem}{Theorem}

\newtheorem{lemma}{Lemma}

\newtheorem*{lemma2}{Lemma 2'}
\newtheorem*{lemma3}{Lemma 3'}

\renewcommand{\S}{\mathcal{S}}
\renewcommand{\P}{\mathcal{P}}
\newcommand{\Q}{\mathcal{Q}}
\newcommand{\T}{\mathcal{T}}
\title{Tight Bounds on the  Average Length, Entropy, and Redundancy of Anti-Uniform Huffman Codes}
\author{\hspace{-3mm}
\begin{tabular}{cc}
Soheil Mohajer & Ali Kakhbod\\ 
\small{ School of Comp. and Comm. Sciences\ \ \ } & \small{Dept. Elec. and Comp. Engineering}\\
\small{EPFL, Switzerland} &  \small{Isfahan University of Technology, Iran}\\
\small{\texttt{soheil.mohajer@epfl.ch}} & \small{\texttt{ali\_kakhbod@ec.iut.ac.ir}}
\end{tabular}
}
\date{}
\begin{document}

\maketitle

\begin{abstract}
In this paper we consider the class of anti-uniform Huffman codes and derive tight lower and upper bounds on the average length, entropy, and redundancy of such codes in terms of the alphabet size of the source. The Fibonacci distributions are introduced which play a fundamental role in AUH codes. It is shown that such distributions maximize the average length and the entropy of the code for a given alphabet size. Another previously known bound on the entropy for given average length follows immediately
from our results.
\end{abstract}

\section{Introduction}

Consider a discrete source with finite size alphabet $\S=\{s_1,s_2,\dots,s_n\}$ and associated ordered probability distribution $\P=(p_1,p_n,\dots,p_n)$ where $p_1\geq p_2\geq\cdots\geq p_n$. It is well-known that the Huffman encoding algorithm \cite{Huffman52} provides an optimal prefix-free code for this source. A binary Huffman code is usually represented using a binary tree $\T$, whose leaves correspond to the source symbols; The two edges emanating from each intermediate node of $\T$ are labeled with either $0$ and $1$, and the codeword corresponding to a symbol is the string of labels on the path from the root to the corresponding leaf.
Huffman's algorithm is a recursive bottom-up construction of $\T$, where at each time the two smallest probabilities are merged 
into a new unit, and henceforth represented by an intermediate node in the tree. 

We denote by $l_i$ the length of the codeword associated to symbol $s_i$ which is the number of edges from the root to the node  $s_i$ on the Huffman tree. 
Then, the expected length of the Huffman code is defined as
\begin{equation}
L(\P)=\sum_{i=1}^n p_i l_i.
\end{equation}
Similarly, the entropy of the source is defined as
\begin{equation}
H(\P)=-\sum_{i=1}^n p_i \log p_i,
\end{equation}
where all the logarithms in this paper are in base $2$. The Huffman encoding is optimal in the sense that no other code for distribution $\P$ can have a smaller expected length than $L(\P)$.
The redundancy $R(\P)$ of the code is defined as the difference between the average codeword length $L(\P)$, and the entropy $H(\P)$ of the source.
It is easy to show that the redundancy of the Huffman code is always non-negative and never exceed $1$.

\begin{figure} 
\centering \input{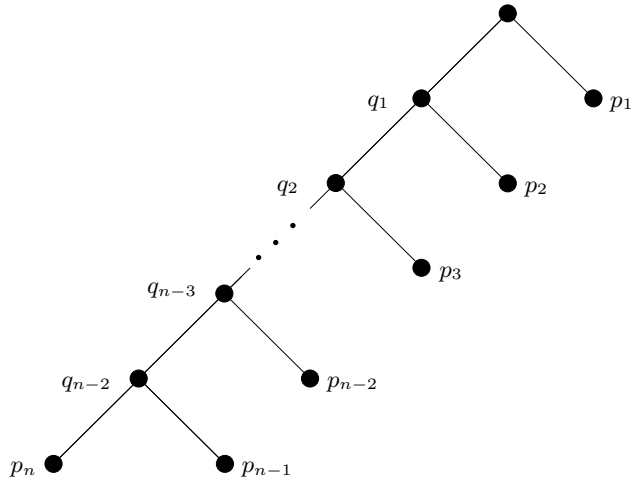}
\caption{An anti-uniform Huffman tree for a source with $n$ symbols.}\label{auh}
\end{figure}

In contrast with uniform Huffman code wherein $|l_i-l_j|\leq 1$, a code (source) is called \emph{anti-uniform Huffman} (AUH) \cite{ali} (Fig.\ref{auh}) if $l_i=i$ for $i=1,\dots,n-1$ and $l_n=n-1$. Such sources can be generated by several probability distributions. It has been shown in \cite{hum} that the normalized \emph{tail} of the Poisson distribution satisfies AUH structure. These kinds of distributions are also considered by Kato \emph{et. al.} \cite{kat} and in particular it is shown that the geometric distribution with success probability greater than some critical value satisfies AUH condition. 

The class of AUH sources are also known for their property of achieving minimum redundancy in different situations. It has been shown in \cite{soheil} that AUH codes potentially achieve the minimum redundancy of Huffman code of a source for which the probability of one of the symbols is known. A similar result by Capocelli \emph{et. al.} \cite{Cap-DeS91} shows that AUH structure achieves the minimum redundancy of a Huffman codes when $p_n$, the probability of the least likely symbol is known. 

In this paper we consider the AUH structure and obtain tight bounds on the average codeword length, entropy and redundancy of such codes in terms of $n$, the alphabet size of the sources. The rest of the paper is organized as follows. We will start with a useful lemma in the next section. Then we state and prove our bound on the average length, entropy and redundancy of AUH codes in Sections \ref{secL}, \ref{secH}, and \ref{secR}, respectively. Finally we conclude in Section \ref{con}. 

\section{Preliminaries}
\begin{figure}
\centering \input{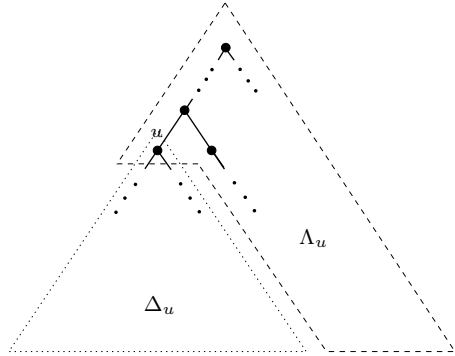}
\caption{Decomposition of a Huffman tree around an intermediate node $u$.}\label{def}
\end{figure}

One can simply define the probability of an intermediate nodes on the Huffman tree as the sum of the probabilities of the leaves lying under it. In an AUH tree of a source with $n$ symbols in Fig.\ref{auh}, there are $n-2$ intermediate nodes which are labeled by $q_1,\dots,q_{n-2}$. We denote the part of a Huffman tree lying under any intermediate node, $u$, by $\Delta_u$ (see Fig.\ref{def}). It is clear that $\Delta_u$ is a subtree which satisfies the Huffman structure, unless the probability of the root is not one. So by normalizing the probabilities of all the leaves by $u$, the probability of the intermediate node, we obtain a new Huffman tree which is denoted by $u^{-1}*\Delta_u$. On the other hand, we can merge all the leaves lying in a subtree $\Delta_u$ in $\T$ and obtain a new Huffman tree which is denoted by $\Lambda_u$. 
The following lemma \cite{soheil} relates the parameters of a source and the its corresponding tree to the parameters of its subtrees.
\begin{lemma}\label{soheil}
For any intermediate node with probability $u$, 
\begin{eqnarray}
H(\T)=H(\Lambda_u)+u H(u^{-1}*\Delta_u).
\end{eqnarray}
The same equation holds holds for average length and redundancy.
\end{lemma}


\section{Average Length}\label{secL}
The average length of any non-trivial code is lower bounded by $1$. Using Lemma. \ref{soheil}, it can be shown that for any arbitrary $n$ the average length of the AUH source with distribution
\[\P_{n,\varepsilon}=(1-\varepsilon,\frac{\varepsilon}{2},\frac{\varepsilon}{4},\dots,\frac{\varepsilon}{2^{n-3}},\frac{\varepsilon}{2^{n-2}},\frac{\varepsilon}{2^{n-2}} )\]
tends to $1$ as $\varepsilon\rightarrow 0$. Therefore, average length of an AUH code is tightly lowerbounded by $1$.

In the following we will state a tight upperbound on the average length of AUH codes in terms of alphabet size of the source. A similar result is also shown independently in the upcoming paper \cite{esm}.

\begin{theorem}\label{L}
Let $\P$ be a distribution over a discrete source of alphabet size $n$. Then $L(\P)$ is upperbounded by
\begin{equation}
L^{\max}_n = \frac{f_{n+3}-3}{f_{n+1}}
\end{equation}
where $f_n$ is the $n$-th Fibonacci number defined as $f_1=f_2=1$ and
\begin{equation}
f_n=f_{n-1}+f_{n-2}\qquad n\geq 3.
\end{equation} 
Furthermore, this bound is tight and can be achieved by the Fibonacci distribution
\[\P^{(F)}_n=\left(\frac{f_{n-1}}{f_{n+1}},\frac{f_{n-2}}{f_{n+1}},\dots,\frac{f_{3}}{f_{n+1}},\frac{f_{2}}{f_{n+1}},\frac{f_{1}}{f_{n+1}},\frac{f_{2}}{f_{n+1}}\right).\]
\end{theorem} 

Before stating the proof, we show two simple lemmas which simplify the proof.  
\begin{lemma}\label{pq}
In any probability distribution $\P=(p_1,p_2,\dots,p_n)$ which maximizes the average length, the probability of any arbitrary leaf is not grater that the probability of the intermediate node in the same level, i.e.,
\[p_i\leq q_i\qquad i=1,\dots,n-2.\]
where $q_i=\sum_{j>i}p_j$.
\end{lemma}
\begin{proof}
Let $p_i>q_i$ for some $i$. This implies $\varepsilon=(p_i-q_i)/2$ is positive. Defining $\varepsilon_i=-\varepsilon$ and $\varepsilon_k=p_k\varepsilon/q_i$ for $k>i$, we can show the distribution 
\[\P'=\left(p_1,\dots,p_{i-1},p_i+\varepsilon_i,p_{i+1}+\varepsilon_{i+1},\dots,p_n+\varepsilon_n\right)\] 
satisfies the AUH constraints and
\begin{eqnarray*}
L(\P')-L(\P)&=& \sum_{k=i}^{n-1}\varepsilon_k k +\varepsilon_n (n-1)\\
&=&\sum_{k=i+1}^{n-1} \varepsilon_k (k-i)+\varepsilon_n (n-1-i)>0
\end{eqnarray*}
which is in contradiction with the maximality of $\P$. 
\end{proof}

\begin{lemma}\label{pq2}
Any probability distribution $\P=(p_1,p_2,\dots,p_n)$ with maximum average length, satisfies 
\begin{eqnarray}
p_1=q_2=\sum_{i>2}p_3.
\end{eqnarray}
\end{lemma}
\begin{proof}
The structure of the Huffman tree and Lemma.\ref{pq} imply $p_1\geq q_2\geq p_2$. Assume the LHS inequality is strict and  so $\varepsilon=(p_1-q_2)/2$ is positive. Defining $\varepsilon_1=-\varepsilon$, $\varepsilon_k=p_k\varepsilon/q_i$ for $k>1$, and
\[\P'=\left(p_1+\varepsilon_1,p_2+\varepsilon_2,\dots,p_n+\varepsilon_n\right),\] 
we have 
\begin{eqnarray*}
L(\P')-L(\P)&=&  \sum_{k=1}^{n-1}\varepsilon_k k +\varepsilon_n (n-1)\\
&=&\sum_{k=2}^{n-1}\varepsilon_k (k-1)+\varepsilon_n (n-2)>0
\end{eqnarray*}
which refuses the maximality of $\P$. 
\end{proof}

\begin{proof}[Proof of Theorem.\ref{L}]
We proof the theorem using induction over the alphabet size, $n$. It is clear that $L^{\max}_2=1=(f_5-3)/f_3$. For $n=3$, one can argue that $\P=(1/3,1/3,1/3)$ has the maximum average length $L^{\max}_3=(f_6-3)/f_4=5/3$. Let the theorem is true for any $k<n$, and $\P=(p_1,p_2,\dots,p_n)$ achieves the maximum average length of an AUH for $n$ symbols. We consider two case as follows.
\begin{description}
\item[(i)] $p_1\geq \frac{f_{n-1}}{f_n+1}$: We denote the subtree lying under $q_1=i-p_1$ by $\Delta_{1-p_1}$ as before. Using Lemma.\ref{soheil}, we have
\begin{eqnarray*}
L(\P)&=&1+(1-p_1)L((1-p_1)^{-1}*\Delta_{1-p_1})\\
&\leq& 1+(1-\frac{f_{n-1}}{f_{n+1}})\frac{f_{n+2}-3}{f_n}\\
&=&\frac{f_{n+3}-3}{f_{n+1}}
\end{eqnarray*}
where the inequality follows from the assumption of the induction for $n=k-1$.

\item[(ii)] $p_1\leq \frac{f_{n-1}}{f_n+1}$: Using Lemma. \ref{pq2} we have $q_2=p_1$. By expanding $L(\P)$ with respect to $q_2$ and using Lemma. \ref{soheil} we can write
\begin{eqnarray*}
L(\P)&=&L(\Lambda_{q_2})+q_2 L(q_2^{-1}*\Delta_{q_2})\\
&=&2+p_1 (L(p_1^{-1}*\Delta_{q_2})-1)\\
&\leq& 2+\frac{f_{n-1}}{f_{n+1}} (\frac{f_{n+1}-3}{f_{n-1}}-1)\\
&=&\frac{f_{n+3}-3}{f_{n+1}}
\end{eqnarray*}
where the inequality is the assumption of the induction for $n=k-2$.
\end{description}
\end{proof}

\noindent\emph{Remark.1 }
One can simply show that $\P^{(F)}$ meets the upper bound for any alphabet size. Although some other distributions such as $\P_4=(0.35,0.30,0.20,0.15)$ can meet the bound, it can be shown that the maximal distribution is unique for $n>4$. 

\noindent\emph{Remark.2 } 
Note that the Fibonacci probability distribution, tends to 
\[\left(t^{2},t^{3},\dots,t^{n-1},t^{n},t^{n-1}\right)\] 
as $n\rightarrow\infty$, where $t=\frac{\sqrt{5}-1}{2}$ is the positive root of $x^2+x-1=0$. Furthermore, It is easy to see that $\{L^{\max}_n\}_{n=1}^\infty$ is an increasing sequence and tends to $t^{-2}=\frac{3+\sqrt{5}}{2}\simeq2.618$ in the asymptotic case.

\section{Entropy}\label{secH}
Since only very particular sources satisfy the AUH structure, the range of the entropy of such sources is not so wide. It is easy to check that the minimum entropy of such sources can be arbitrary close to zero for any alphabet size $n$. In order to see this, one may compute the entropy of 
\[P_{n,\varepsilon}=(1-\varepsilon,\frac{\varepsilon}{2},\frac{\varepsilon}{4},\dots,\frac{\varepsilon}{2^{n-3}},\frac{\varepsilon}{2^{n-2}},\frac{\varepsilon}{2^{n-2}} )\]
for $\varepsilon\leq 2/3$ and show $H(P_{n,\varepsilon})\rightarrow 0$ as $\varepsilon\rightarrow0$.

In spite of that, upperbounding the entropy of AUH codes is not trivial. 
It has been shown in \cite{ali} that the entropy of an \emph{infinite length} AUH codes with given average length  $L$ is upper bounded by 
\begin{eqnarray}\label{ali-L}
H^{\max}_\infty(L)=L\log L -(L-1)\log (L-1). 
\end{eqnarray}
This bound is only valid for infinite source. It can be shown that any dyadic source ($p=2^{-l}$ for some integer $l$) has entropy larger than $H^{\max}_\infty(L)$. The following theorem states a tight upperbound on the entropy of AUH sources with $n$ symbols.

\begin{theorem}\label{H}
The entropy of finite source with $n$ symbols is upperbounded by 
\begin{eqnarray}
H^{\max}_n=H(\P^{(F)}_n)=\log f_{n+1}-\frac{1}{f_{n+1}}\sum_{i=1}^{n-1} f_i\log f_i.
\label{Hmax}
\end{eqnarray}
\end{theorem}

The proof of this theorem is fairly similar to that of Theorem.\ref{L}. The following Lemmas show some basic properties on distributions which achieve the maximum entropy.
 
\begin{lemma2}
Let $\P=(p_1,p_2,\dots,p_n)$ be a distribution over $n$ symbols with maximum Entropy. Then $p_i\leq q_i$for any $i=1,\dots,n-2$.
\end{lemma2}
\begin{proof}
Similar for proof of Lemma.\ref{pq}, we assume that the condition does not hold for some $i$ and make a contradiction. 
Assume $\varepsilon=(p_i-q_i)/2>0$, and define the modified distribution
\begin{eqnarray*}
\P'=(p'_1,\dots,p'_n)=\left(p_1,p_2,\dots,p_{i-1},p_i+\varepsilon_i,p_{i+1}+\varepsilon_{i+1},\dots,p_{n}+\varepsilon_{n}\right),
\end{eqnarray*}
where $\varepsilon_i=-\varepsilon$ and $\varepsilon_{k}=p_k\varepsilon/q_i$ for $k>i$. We can write
\begin{eqnarray*}
H(\P')-H(\P)&=&\sum_k p_k \log p_k -\sum_k p'_k\log p'_k \\
&=&\sum_k p_k \log p_k -\sum_{k<i} p_k\log p'_k -\sum_{k\geq i}(p_k+\varepsilon_k) \log p'_k\\
&=&D(\P||\P') +\sum_{k>i}\varepsilon_k \log \frac{p'_i}{p'_k}>0
 \end{eqnarray*}
where $D(\cdot||\cdot)$ is the Kullback-Leibler divergence and the last inequality follows from the facts that $\sum_{k<i}\varepsilon_k=-\varepsilon_i$ and $\P'$ is an decreasing sequence. This inequality is in contradiction with assumption, which implies the desired result.
\end{proof}
 
\begin{lemma3}
For any distribution  $\P=(p_1,p_2,\dots,p_n)$ which achieves the maximum entropy, $p_1=q_2$.  
\end{lemma3}
\begin{proof}
The structure of AUH tree implies $p_1\geq q_2$. If the inequality is strict, by Lemma.2' we have, $p_1>q_2\geq p_2$. Define $\varepsilon=(p_1-q_2)/2>0$, and consider the distribution 
\[\P'=(p'_1,\dots,p'_n)=\left(p_1+\varepsilon_1,p_2+\varepsilon_2,\dots,p_{n}+\varepsilon_{n}\right),\] 
where $\varepsilon_1=-\varepsilon$ and $\varepsilon_{k}=p_k\varepsilon/q_1$ for $k>1$. We have
\begin{eqnarray*}
H(\P')-H(\P)&=&\sum_k p_k \log p_k -\sum_k p'_k\log p'_k \\
&=&\sum_k p_k \log p_k -\sum_{k} p'_k\log (p_k+\varepsilon_k \\
&=&D(\P||\P') +\sum_{k>1}\varepsilon_k \log \frac{p'_k}{p'_i}>0
\end{eqnarray*}
which refuses the assumption we made in the beginning. 
\end{proof}

\begin{proof}[Proof of Theorem.\ref{H}]
Similar to the proof of Theorem.\ref{L}, we prove this theorem by induction over the alphabet size of the source. Since the uniform distribution satisfies the AUH constraints for $n=3$ and $n=2$, we have $L^{\max}_2=1$ and $L^{\max}_3=\log3$ which coincide with (\ref{Hmax}). Now, let $n\geq 4$ and the bound is valid for $k<n$. We consider two cases. 
\begin{description}
\item [(i)] $p\geq\frac{f_{n-1}}{f_{n+1}}$: Using Lemma. \ref{soheil} and by expansion the entropy with respect to $q_1$, we have
\begin{eqnarray}
H(\P)&=&h(p_1)+ q_1 H(q_1^{-1}*\Delta_{q_1})\nonumber\\
&\leq&h(p_1)+ (1-p_1)H^{\max}_{n-1}\nonumber\\
&\leq&h(\frac{f_{n-1}}{f_{n+1}})+(1-\frac{f_{n-1}}{f_{n+1}})H^{\max}_{n-1}\nonumber\\
&=&\log f_{n+1}-\frac{1}{f_{n+1}}\sum_{i=1}^{n-1} f_i\log f_i
\end{eqnarray}
where  the first inequality comes form the assumption of induction for $k=n-1$ and the second inequality follows from the fact that the function $\alpha(x)=h(x)+(1-x)H^{\max}_{n-1}$ is non-increasing for $x\geq f_{n-1}/f_{n+1}$. It can shown by the taking the derivative of $\alpha(x)$ as
\begin{eqnarray*}
\frac{d\alpha}{dx} &=&\log \frac{1-p}{p}-H^{\max}_{n-1} \ \leq \ \log \frac{1-f_{n-1}/f_{n+1}}{f_{n-1}/f_{n+1}}-H^{\max}_{n-1}\\
&=&\log \frac{f_n}{f_{n-1}}-\log f_{n}+\frac{1}{f_{n}}\sum_{i=1}^{n-2} f_i\log f_i\\
&=&\frac{1}{f_{n}}\left[\sum_{i=1}^{n-2} f_i\log \frac{f_i}{f_{n-1}}-\log f_{n-1}\right]\leq 0.
\end{eqnarray*}

\item [(ii)] $p\leq\frac{f_{n-1}}{f_{n+1}}$: Using Lemma. 3', we can only focus on the distributions for which $q_2=p_1$. Now we use Lemma. \ref{soheil} to expand the entropy with respect to  $q_2$.
\begin{eqnarray}
H(\P)&=&h(\Lambda_{q_2})+ q_2 H(q_2^{-1}*\Delta_{q_2})\nonumber\\
&=&-2p\log p -(1-2p)\log(1-2p) +p H(q_2^{-1}*\Delta_{q_2})\nonumber\\
&\leq& -2p\log p -(1-2p)\log(1-2p) +p H^{\max}_{n-2}\nonumber\\
&\leq& -2\frac{f_{n-1}}{f_{n+1}}\log \frac{f_{n-1}}{f_{n+1}}-(1-2\frac{f_{n-1}}{f_{n+1}})\log\left(1-2\frac{f_{n-1}}{f_{n+1}}\right) +\frac{f_{n-1}}{f_{n+1}} H^{\max}_{n-2}\nonumber\\
&=&\log f_{n+1}-\frac{1}{f_{n+1}}\sum_{i=1}^{n-1} f_i\log f_i
\end{eqnarray}
where again the assumption of induction for $k=n-2$ implies the first inequality and the second one follows from the fact that  $\beta(x)=-2x\log x -(1-2x)\log(1-2x) +x H^{\max}_{n-2}$ is an increasing function for $x>f_{n-1}/f_{n+1}$. 
\end{description}
%
\end{proof}

\noindent\emph{Corollary.} Note that the maximum achievable entropy for an infinite size alphabet is
\begin{eqnarray}
\lim_{n\rightarrow\infty} H^{\max}_n=- \sum_{i=2}^{\infty} t^{i}\log t^{i}=\left(1+\frac{1}{t^2}\right)\log\frac{1}{t}
\end{eqnarray}
which is obtained for $\P^{(F)}_\infty$ and coincides with 
\begin{eqnarray*}
L^{\max}_{\infty}\log L^{\max}_{\infty}-(L^{\max}_{\infty}-1)\log(L^{\max}_{\infty}-1)\hspace{-5pt}&=&\hspace{-5pt}\frac{1}{t^2}\log \frac{1}{t^2}-(\frac{1}{t^2}-1)\log(\frac{1}{t^2}-1)\\
&=&\hspace{-5pt}\left(1+\frac{1}{t^2}\right)\log\frac{1}{t}.
\end{eqnarray*}
This simply proves (\ref{ali-L}).

\section{Redundancy}\label{secR}
It is known that the Huffman code associated to any dyadic source has zero redundancy and since such distributions exist for any arbitrary alphabet size and satisfy the AUH constraints, the redundancy of AUH code is tightly lowerbounded by zero. On the other extreme, it can be shown that the redundancy of AUH codes can be arbitrary close to $1$. 

\begin{theorem}
For and alphabet size $n$ and $\delta>0$, there exist probability distributions on $n$ symbols for which 
\[R(\P)>1-\delta.\]
\end{theorem}
\begin{proof}
Take an arbitrary AUH distribution $\Q=(q_1,\dots,q_{n-1})$ over $n-1$ symbols and $\varepsilon>0$ small enough such that $1-\varepsilon\geq\max\{q_1\varepsilon,(1-q_1)\varepsilon\}$. Therefore, $\P=(1-\varepsilon,q_1\epsilon,\dots,q_{n-1}\varepsilon)$ is a distribution with AUH code. Using Lemma.\ref{soheil} we have
\begin{eqnarray*}
R(\P)&=&1-h(\varepsilon)+\varepsilon R(\varepsilon^{-1}*\Delta_\varepsilon)\\
&=& 1-h(\varepsilon)+\varepsilon R(Q)
\end{eqnarray*}
where $h(x)=-x\log x -(1-x)\log(1-x)$ is the binary entropy function. Note that $R(Q)$ is bounded by $L(Q)\leq L^{\max}_{n-1}<t^{-2}$, and $R(\P)$ tends to $1$ as $\varepsilon\rightarrow 0$.
\end{proof}

\section{Conclusion}\label{con}
In this paper we have obtained tight upper and lower bounds on the average length, entropy, and redundancy of the Huffman code for an anti-uniform Huffman source. We showed that for a given alphabet size, Fibonacci distributions maximize the average length and entropy.

\bibliographystyle{IEEEtran}

\end{document}